\documentclass[twocolumn,amsmath,amssymb]{revtex4}

\usepackage{graphicx}
\usepackage{epstopdf}
\usepackage{dcolumn}
\usepackage{amsmath}
\usepackage{subfigure}

\begin{document}

\renewcommand{\thefigure}{\arabic{figure}}

\title{Measuring chemical composition and particle cross-section of ultra-high energy cosmic rays by a ground radio array}

\author{
 K.~Belov\\
University of California, Los Angeles\\
UCLA Department of Physics and Astronomy / 154705\\
475 Portola Plaza\\
Los Angeles, CA 90095, USA} 

\date{\today}

\begin{abstract}
We present a technique to measure chemical composition and particle cross-section of ultra-high energy cosmic rays using radio data. We relate the geometry of the radio footprint on the ground to the depth of the extensive air shower maximum, $X_{max}$. We suggest to use the spectral information of the radio signal to improve the $X_{max}$ reconstruction by minimum number of antennas on the ground.

\vspace{1pc}
\end{abstract}

\maketitle

\section{Introduction}
The very existence of ultra-high energy cosmic rays, UHECRs, presents a scientific challenge. Known mechanisms struggle to explain acceleration of charged particles up to $10^{20}$ eV, the energies observed by cosmic ray detectors \cite{PhysRevLett.71.3401, Abbasi200953, Abraham2010239}. UHECRs are considered to be nonelectromagnetic messengers from the Universe that can bring information about their sources, acceleration mechanisms, interstellar medium they propagate through and potentially indicate new physics. While there is a good agreement about the UHECR energy spectrum among different cosmic-ray experiments \cite{Abbasi200953, AbuZayyad201316, Abraham2010239}, the cosmic-ray chemical composition at the end of the spectrum is still unknown. At the same time, mass composition is the key for understanding the origin and acceleration mechanisms of the UHECRs and must be known in order to measure properties of particle interactions at the energies, exceeding the capabilities of modern accelerators. We are proposing a new technique to measure the chemical composition and inelastic cross-section of the UHECRs using radio observations.

There are outstanding disagreements on UHECR chemical composition. There are some indications that the composition is predominantly light \cite{PhysRevLett.104.161101, Matthews201179, Glushkov2013}, but other experimental data suggest the opposite \cite{PhysRevLett.104.091101}. The sources of the UHECRs are yet to be identified. The major cosmic-ray experiments reported mixed results on the source anisotropy. The central value of the correlation fraction with nearby Active Galactic Nuclei, AGNs, from Veron-Cetty Catalog, VCV, as possible candidates, initially reported by Pierre Auger experiment in southern hemisphere~\cite{Abraham2008188}, has decreased as more data became available~\cite{Abreu2010314, AugerAGN2011_1475-7516-2012-04-040, AugerAnisotropyICRC2013}. However, the significance level remains unchanged. At the same time, the Telescope Array experiment in northern hemisphere reported no significant correlations with nearby AGNs \cite{TA_Anysotropy2012_0004-637X-757-1-26} as did its predecessor, the HiRes experiment~\cite{Abbasi2008175}.
Due to the magnetic field deflection, the controversy of the former result with the AGN correlation by the same experiment~\cite{Abraham2008188, Abreu2010314, AugerAGN2011_1475-7516-2012-04-040, AugerAnisotropyICRC2013} can be explained by an excess flux from the galactic center~\cite{PhysRevLett.105.091101}. However, there is no data confirming the excess yet. Identification of the UHECR sources relies on the mass composition measurements.

\begin{figure}[h]
\includegraphics[width=\columnwidth]{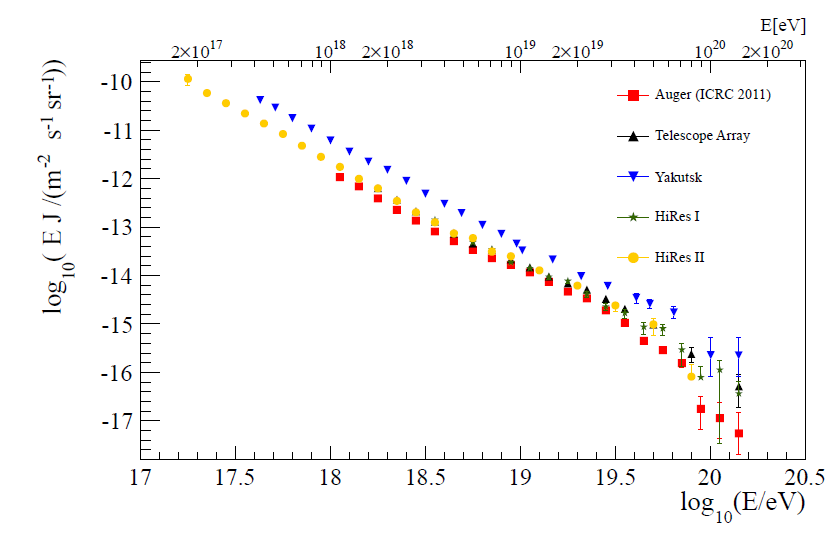}
\caption{Published energy spectra, with the flux multiplied by the primary particle energy, for Auger (combined Hybrid/SD) \cite{Salamida:ICRC2013}, TA SD \cite{2041-8205-768-1-L1},
Yakutsk SD \cite{Yakutsk:2010}, HiRes~I \cite{PhysRevLett.100.101101}, and HiRes~II \cite{PhysRevLett.100.101101}. Adopted from~\cite{AugerTAYakutsjSpectrum:2013}.}
\label{AugerTAYakutskSpectrum2012:fig} 
\end{figure}

The Greisen--Zatsepin--Kuzmin, GZK, feature~\cite{greisen:gzk66}, possibly observed by some cosmic-ray experiments \cite{PhysRevLett.100.101101, Abraham2010239, Yakutsk:2010} as a suppression at the higher end of the energy spectrum, see figure \ref{AugerTAYakutskSpectrum2012:fig}, 
limits the distance of travel for the highest energy cosmic rays by invoking a mechanism of photo-disintegration of protons above 10$^{19.6}$~eV. The predicted GZK ``event horizon'' for protons is 50--100~Mpc and is even shorter for heavier nuclei.  Despite the lack of suitable sources within this distance, many higher energy particles were detected. If real, the GZK mechanism should provide a flux of ultra-high energy neutrinos~\cite{BerezinskyZatsepin1970}. However, two ANITA balloon--borne experiments in Antarctica did not discover the neutrino flux above the expected background~\cite{Gorham200910, PhysRevD.82.022004}. If UHECR composition is predominantly light, the lack of the GZK neutrinos can indicate the Lorentz invariance violation~\cite{Coleman1997249, Scully2011575}, providing a hint for physics beyond the Standard Model. 

The suppression at the end of the cosmic-ray spectrum can also be explained if heavy nuclei contribution to the cosmic-ray flux increases with energy. In this scenario, heavy nuclei could be accelerated by powerful sources that existed in our Galaxy long time ago, such as GRBs, hypernovae, collapsars or other unusual supernova explosions. If the galactic magnetic fields are stronger than it is assumed today, these heavy nuclei can have a long diffusion time, providing an isotropic, composition-dependent flux that agrees with the current observations~\cite{PhysRevLett.105.091101}.

\begin{figure}[h]
\includegraphics[width=\columnwidth]{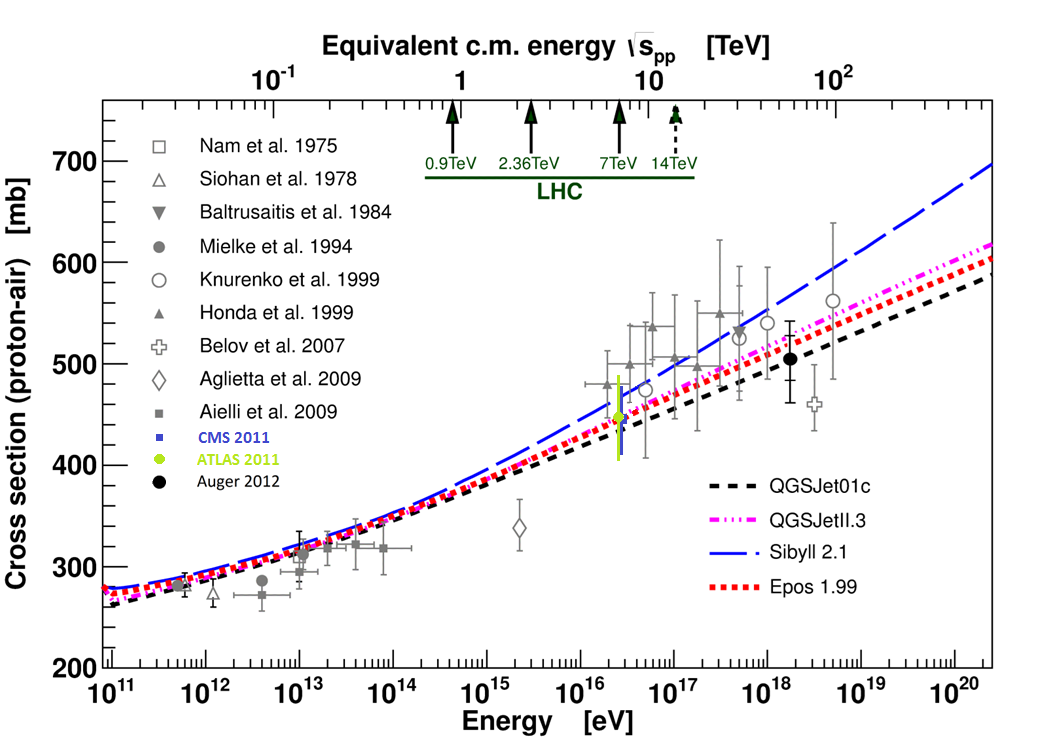}
\caption{Proton-air  inelastic cross-section measured by accelerators and cosmic-ray experiments. Color lines show different interaction models and accelerator data extrapolation.}
\label{p-air_inelastic_cross-section:fig} 
\end{figure}

To date, the cosmic rays are the only source of particles to study interactions at extreme energies \cite{Belov2006197, PhysRevLett.109.062002}. Figure \ref{p-air_inelastic_cross-section:fig} shows the proton-air inelastic cross-section measurements by accelerators and cosmic-ray experiments.
It should be noted that changing particle interaction properties at these energies can alter the development of extensive air showers, EASs, cascades of secondary particles in the earth's atmosphere observed to study the UHECRs. This change in the EAS development can be misinterpreted as the change in chemical composition.

The low cosmic-ray flux at ultra-high energies makes composition and cross-section measurements extremely difficult by limiting the available data. We propose a new measurement technique utilizing radio emission from EASs to significantly increase the amount of UHECR data and to complement and cross-calibrate the measurements done by existing air fluorescence and ground counter UHECR observatories. Probabilistic composition-tagging of each event will allow us to backtrack the proton-like cosmic-ray events with smaller deflection in the galactic magnetic field helping the source identification. A precise measurement of the composition change with the energy will reveal the nature of the spectral cutoff at the extreme energies.

\section{UHECR composition and cross-section measurements}


Direct measurements of the cosmic ray chemical composition are only possible for the cosmic rays with energies below $10^{14}$ eV. The cosmic-ray flux diminishes at ultra-high energies, requiring an adoption of indirect measurement techniques. One of the most reliable indirect techniques is the observation of ultra-violet, UV, air fluorescence light caused by EASs. The air fluorescence can be observed from the ground or from space \cite{PhysRevLett.39.847, Abraham2010227, TakahashiJEM-EUSO:2009}. The total amount of light emitted along the extensive air shower is proportional to the number of charged particles in the cascade, the air shower profile. The number of charged particles in the cascade reaches the maximum when ionization losses equal bremsstrahlung and pair production, and all particles reach the critical energy of about 80 MeV. The depth of the shower maximum is usually refereed as $X_{max}$ and is measured in g/cm$^2$. The mean $X_{max}$ and the shape of the $X_{max }$ distribution at different primary particle energies can be related to the chemical composition and particle cross-section by comparing to theoretical models~\cite{PhysRevLett.84.4276, PhysRevLett.104.161101, Matthews201179, PhysRevLett.104.091101}. Depending on the EAS geometry, the shower maximum can be in the field of view of an air fluorescence detector proving an accurate $X_{max}$ measurement. Air fluorescence detectors, however, come at a high cost and can only operate during moonless nights, limiting the detector duty cycle by 10\%.


An interaction of a heavy nucleus with the atmospheric nuclei results in higher multiplicity compared to a lighter nucleus interaction. Because energy is shared among nuclear fragments, the resultant secondary pions end up less energetic and have a greater chance to decay into muons, than more energetic pions, produced at earlier stages of the shower development or from a collision by a lighter nucleus. Thus, we are able to observe a difference in the muon content between heavier and a lighter nuclei interactions in the atmosphere and can measure the chemical composition of the primary particles by either measuring the total number of muons in EAS on the ground or by measuring the local density of the muons. However, the statistical muon density fluctuations can be larger than fluctuations between showers caused by different nuclei. These measurements are also model dependent. In addition, the difference in the muon content should also reduce with energy as the path available for the decay of the high energy pions decreases with the shower developing deeper in the atmosphere, limiting the power to resolve differences in chemical composition at ultra-high energies.


The radio technique relies on reconstruction of the radio frequency, RF, emission from an EAS map on the ground. Although, it is also dependent on the RF emission model, it can complement the existing methods by providing data by utilizing a different measurement technique. The RF emission models are derived from first principles \cite{Huege:2013, AlvarezMuniz2012325} by using well known and experimentally confirmed electrodynamics. Radio detectors can operate with 100\% duty cycle and are insensitive to atmospheric conditions, avoiding the known limitations of the air fluorescence detectors. The cost of deployment and operation of a radio detector is also lower, compared to air fluorescence or ground counter arrays. 

\section{Radio emission from extensive air showers}


An EAS is triggered in earth's atmosphere when a high energy cosmic particle strikes. As the cascade of charged secondary particles moves in the atmosphere, a net negative charge is built up due to the Askaryan effect \cite{askaryan:65, PhysRevLett.86.2802}. The excess charge propagation in the atmosphere leads to a radially polarized RF emission component of the electric field. In addition, a deflection of electrons and positrons in the cascade by the earth's magnetic field leads to a transverse current, and, thus, to the RF emission component with the electric field vector directed along the transverse current. The relative contribution of each component for an observer on the ground depends on the local geomagnetic field geometry and location of the observer relative to the shower axis.

Due to a low index of refraction in the air and relativistic amplification of the signal in the Cherenkov angle, the RF emission  power is boosted within a narrow cone in the forward direction of the EAS. Thus, the radio signal from an EAS appears to be concentrated at a Cherenkov cone around the shower axis. For an observer located at the Cherenkov angle, the air shower develops simultaneously and all the emission sums up coherently, which results in a boosted radio impulse. 
The RF emission from an EAS creates an elliptical ``footprint'' on the ground. While the ellipticity of the footprint depends on the air shower zenith angle, the size of the ellipse depends on the depth of the shower maximum, $X_{max}$, the depth in the atmosphere where the density of the charged particles reaches its maximum. 

On the Cherenkov cone, the emission is coherent up to wavelengths comparable to the shower size, which corresponds to GHz in frequency.  Away from the Cherenkov angle the coherence at higher frequencies disappears first, resulting in a frequency spectrum roll off and a loss of the total power. The angular dependence of the RF spectrum can be used to assist in reconstruction of the air shower geometry and the energy of the primary particle by pinpointing the observer location relative to the Cherenkov cone \cite{belovarena20121}.

If the RF measurement is done on the inside of the Cherenkov cone, the RF signal in the time domain is reversed. While the shower is propagating in the atmosphere with the speed $c$, the RF signal is propagating with the speed of light in the medium, $\frac{c}{n}$, where $n$ is the index of refraction. Due to a slower speed of light in the medium, the signal from the end of air shower reaches the observer before the signal from the beginning of the shower, leading to the RF pulse reversal in the time domain. This property can be used to detect the side of the Cherenkov cone the receiving antenna is located at,  increasing the accuracy of the RF mapping and the air shower geometry and profile reconstruction.

\section{MC simulations}

\begin{figure}[t]
\centering
\subfigure[$X_{max}=755~g/cm^2$]{
\includegraphics[scale=0.3]{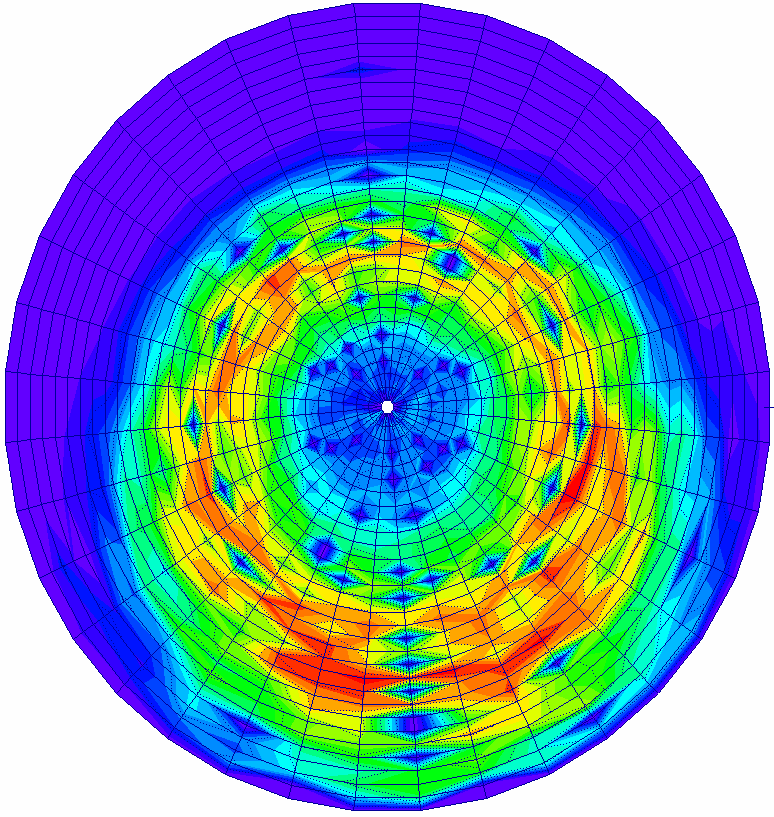}
}
\subfigure[$X_{max}=819~g/cm^2$]{
\includegraphics[scale=0.3]{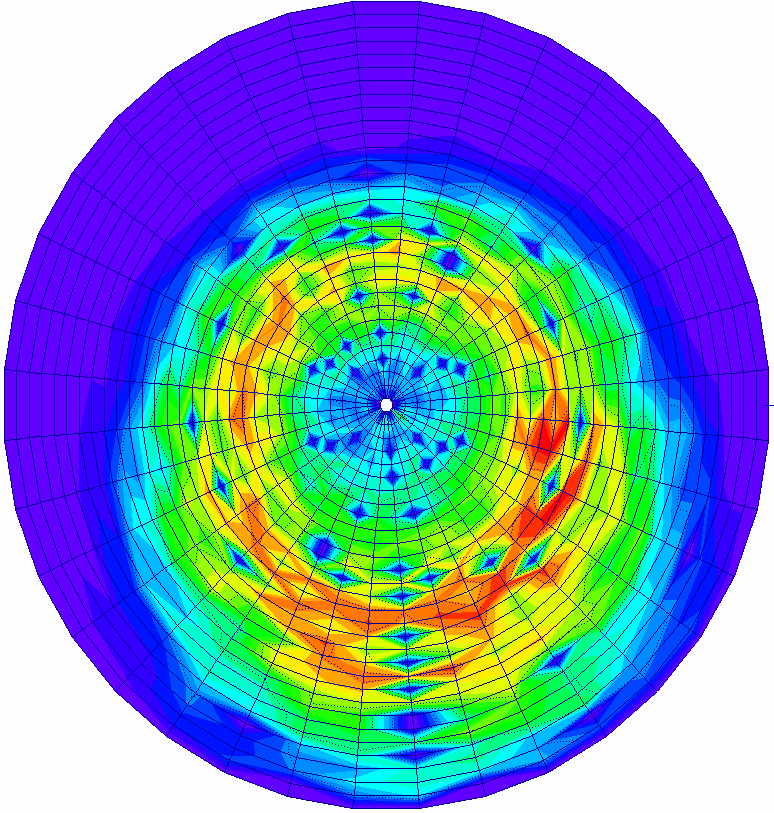}
}
\caption{Simulated radio footprint on the ground for a 55$^\circ$ zenith angle air shower. The circle diameter is 600 m.}
\label{RFfootpring55deg:fig}
\end{figure}

\begin{figure}[t]
\centering
\subfigure[$X_{max}=711~g/cm^2$]{
\includegraphics[scale=0.3]{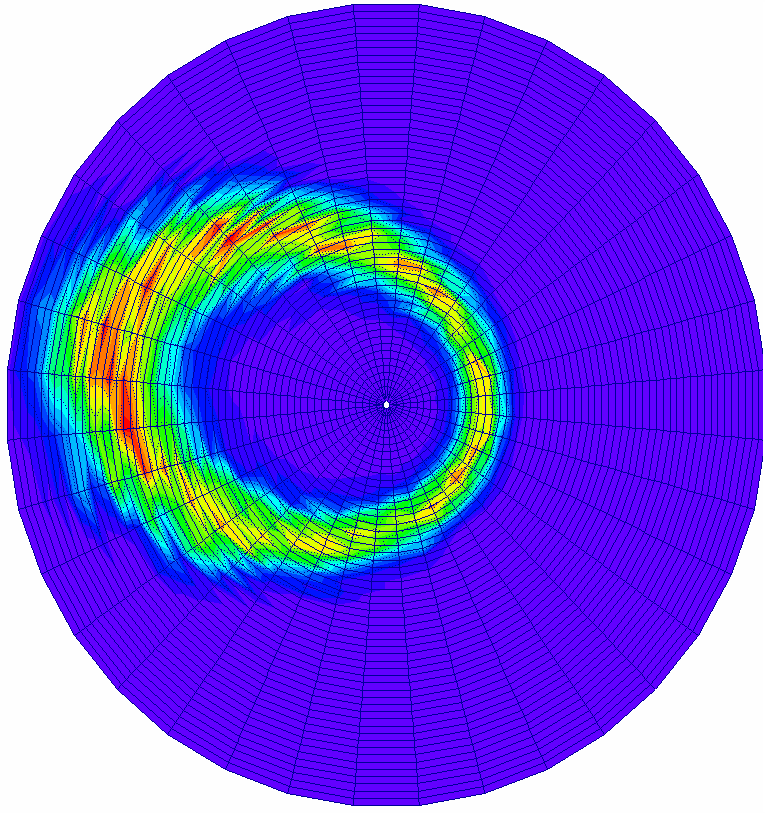}
}
\subfigure[$X_{max}=879~g/cm^2$]{
\includegraphics[scale=0.3]{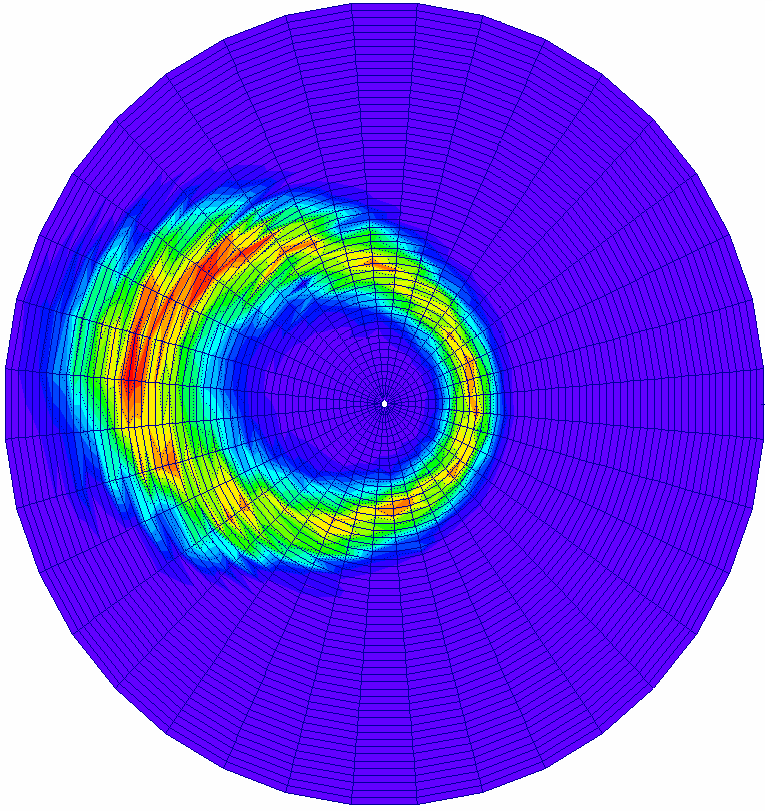}
}
\caption{Simulated radio footprint on the ground for a 70$^\circ$ zenith angle air shower. The circle diameter is 2200 m.}
\label{RFfootprint70deg:fig}
\end{figure}

For our model, we use Monte Carlo, MC, simulations of EAS based on CORSIKA v.6.960~\cite{CORSIKA:98} code and QGSJET01 strong interaction model~\cite{Kalmykov199717}.  The chosen atmospheric model corresponds to an average January in Antarctica. The index of refraction changes with altitude along the shower development path and is driven by our choice of the atmospheric model. The RF emission from the air showers is simulated by CoREAS plug-in~\cite{Huege:2013} to the CORSIKA code. An ensemble of UHECR events with the same geometry and different zenith angles was simulated at~$5\times10^{19}$~eV energy. Examples of simulated radio maps for the air showers with 55$^\circ$ and 70$^\circ$ zenith angles are shown in Figures~\ref{RFfootpring55deg:fig} and~\ref{RFfootprint70deg:fig} correspondingly.

We found the shape of the RF footprint on the ground to be energy independent, as expected. Air showers with different zenith angles have different eccentricity of the ellipse and, thus, different area confined by the ellipse. 

\begin{figure}[htb]
\includegraphics[width=0.8\columnwidth]{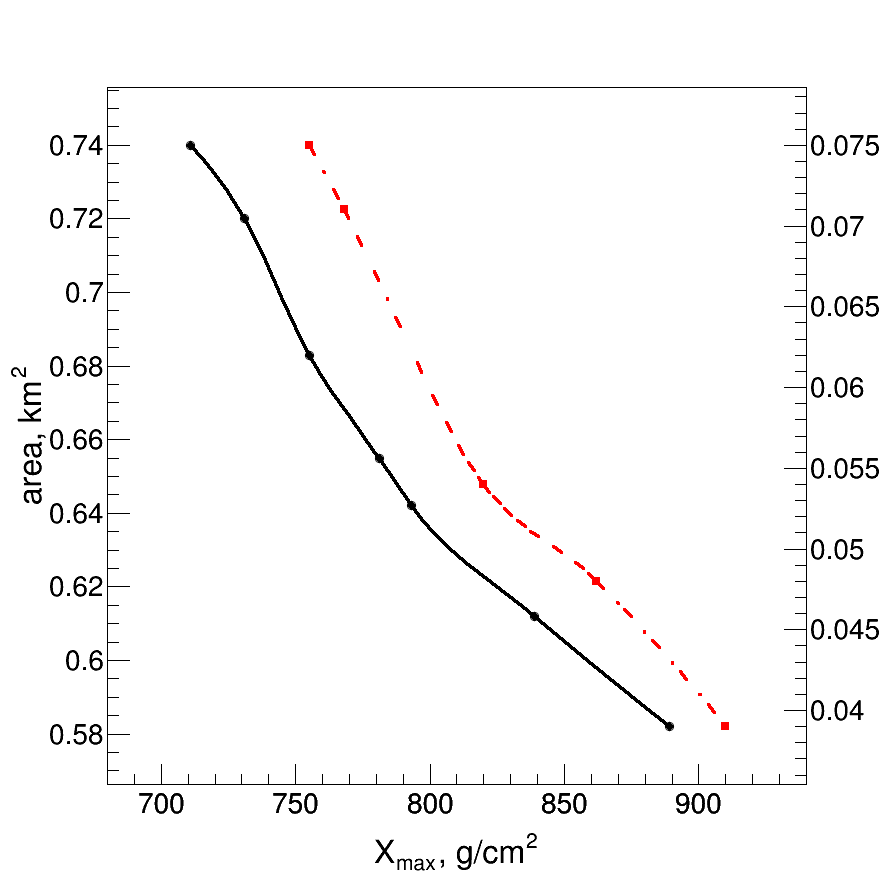}
\caption{Cherenkov ellipse area as a function of the $X_{max}$ for 70$^\circ$ zenith angle shower (left scale, solid line) and 55$^\circ$ zenith angle shower (right scale, dot--dashed line). MC simulation.}
\label{cherenkov_ellipse_area_vs_xmax:fig} 
\end{figure}

The shower zenith angle can be measured from the eccentricity of the reconstructed RF footprint and for the known zenith angle, the area of the ellipse can be related to the EAS profile. The area confined by the Cherenkov ring on the ground, as a function of the $X_{max}$ for 55$^\circ$ and 70$^\circ$ air showers is shown in Figure~\ref{cherenkov_ellipse_area_vs_xmax:fig}.
Reconstruction of the RF map on the ground yields the $X_{max}$, and can be used for the chemical composition and particle cross-section measurements at ultra-high energies. The geometry of the RF footprint also allows us to obtain the primary particle arrival direction.

\section{Discussion}

The geometry of the footprint created by radio emission from EAS on the ground depends on the extensive air shower zenith angle and the shower development in the earth's atmosphere and can be related to the depth of the shower maximum, $X_{max}$. While the RF emission intensity depends on the primary energy of the cosmic particle, the geometry of the RF footprint is energy independent, excluding the $X_{max}$ dependence of the primary particle energy, that can be easily deconvoluted. This opens a possibility for a precise measurement of the chemical composition and particle cross-section at ultra-high energies using an array of radio antennas on the ground.

The $X_{max}$ resolution that can be achieved depends on the accuracy of the RF footprint reconstruction and improves with increasing number of the triggered ground antennas. Although the overall geometry of the RF footprint is energy independent, the edges of detectability expand for a given antenna sensitivity and the ambient electromagnetic interference of anthropogenic nature and cosmological sources such as the galactic center, the sun and transients. The background signals, like radio emission from GRBs and other transients, can contain useful information and can be of a great interest for astronomers and should be considered as a useful signal during the planning of the radio cosmic-ray observatory.

For a 70$^\circ$ zenith angle EAS is the total area of the RF footprint is about 0.6--0.75~km$^2$.  Few antennas have to register the signal in order to accurately reconstruct the RF footprint. The footprint area increases for more inclined showers. Detailed MC simulations have to be done to estimate the number and the density of antennas required to achieve good $X_{max}$ resolution and meaningful statistics at highest energies, but preliminary calculations show that several radio antennas with about 1~km separation are sufficient to detect 1000's of UHECR events above 10$^{18}$~eV.

One way to reduce the number of antennas required, is to measure the frequency of the RF emission in a broad spectrum~\cite{belovarena20121}, which allows pinpointing the location of the triggered antenna relative to the Cherenkov cone. The polarity of the RF signal can eliminate the ambiguity of the of the antenna location on the inside or outside of the Cherenkov cone, improving the reconstruction accuracy. If the antennas are sensitive to both vertical and horizontal polarizations, the radial ambiguity can be resolved, because the local geomagnetic field is known and relative contribution of the horizontal and vertical polarization component of the electric field  can be calculated for a given EAS geometry. The reconstruction of the RF footprint will allow us to point back to the source of the primary cosmic particle. The pointing accuracy that can be achieved ranges from $\sim 2^\circ$, limited by the Cherenkov angle in case of only one antenna hit, to $\sim 0.1\text{--}0.2^\circ$, in case of the full reconstruction of the Cherenkov cone on the ground. This creates new opportunities for charged particle astronomy.

\section{Acknowledgments}
We would like to thank Tim Huege and Marianne Ludwig from Karlsruhe Institute of Technology  for helpful discussions and providing the simulation code used in this work. This material is based upon work supported by the Department of Energy under Award Number DE-SC0009937 and NASA Award Number NNX11AC46G.
\bibliographystyle{KBelovCompositionCrossSectionRadioArray}
\bibliography{C:/UCLA/ANITA/KBelovBibTexRefDatabase}

\end{document}